\documentclass[amsmath,10pt,twocolumn,superscriptaddress,footinbib,pre]{revtex4-2}
\usepackage{amsmath,amsfonts,bm,dsfont}
\usepackage{graphicx}
\usepackage{subfigure}
\usepackage{float}
\usepackage{chemformula}
\usepackage[normalem]{ulem}
\usepackage{physics}
\usepackage{extpfeil}
\usepackage{uri}
\newcommand{\appropto}{\mathrel{\vcenter{
  \offinterlineskip\halign{\hfil$##$\cr
    \propto\cr\noalign{\kern2pt}\sim\cr\noalign{\kern-2pt}}}}}

\usepackage{color}
\usepackage[colorlinks=true,allcolors=blue]{hyperref}
\usepackage{tikz}
\usepackage{upgreek}
\usepackage{amssymb}

\begin{document}

\title{It Takes Two to Make a Thing Go Right: Boosting Current in Coupled Motors}
\author{Geyao Gu}
\author{Drew Alvarez}
\author{John Strahan} 
\affiliation{Department of Chemistry, Northwestern University, 2145 Sheridan Road, Evanston, Illinois 60208, USA}
\author{Alex Albaugh}
\affiliation{Department of Chemical Engineering and Materials Science, Wayne State University, 5050 Anthony Wayne Drive, Detroit, Michigan 48202, USA}
\affiliation{NSF-Simons National Institute for Theory and Mathematics in Biology, Chicago, Illinois 60611, USA}
\author{Emanuele Penocchio}
\email{emanuele.penocchio@northwestern.edu.}
\affiliation{Department of Chemistry, Northwestern University, 2145 Sheridan Road, Evanston, Illinois 60208, USA}
\author{Todd R.~Gingrich}
\email{todd.gingrich@northwestern.edu.}
\affiliation{Department of Chemistry, Northwestern University, 2145 Sheridan Road, Evanston, Illinois 60208, USA}
\affiliation{NSF-Simons National Institute for Theory and Mathematics in Biology, Chicago, Illinois 60611, USA}

\begin{abstract}
  Catalysis-driven synthetic molecular motors operate in a loose mechanochemical coupling regime, one in which a decomposition of a fuel molecule does not reliably produce a forward step.
  In that regime, stochastic backward steps can significantly degrade the motor's current, prompting us to ask whether mechanically coupling multiple such motors can boost their averaged current.
  By simulating rotaxane-based motors with two classes of models\textemdash particle-based nonequilibrium molecular dynamics and jump-diffusion models\textemdash we show that current boosts are physically achievable.
  Our observed boosts, which amplify current by single-digit factors, emerge when coupling between motors can increase the activity, speeding up the rate of both forward and backward steps.
  In doing so, the bias for preferring forward steps actually degrades, but the lost bias can be largely recovered by raising the fuel concentration, demonstrating a general design strategy: amplify activity through coupling and restore bias through stronger driving.
\end{abstract}
\maketitle
\section*{Introduction}

Molecular motors catalyze the decomposition of chemical fuels to generate directional motion~\cite{Brown_2019,Iino_2020,amano2021chemical}, which has the potential to perform work.
Typically, the motors operate in a regime with a surplus of fuel and a dearth of waste, such that the chemical reaction is strongly directional from fuel to waste.
This chemical gradient translates into a strongly biased motor when the mechanochemical coupling is tight~\cite{hirokawa2009kinesin,mallik2004molecular}.
However, a motor with loose mechanochemical coupling can take many stochastic ``backward'' steps even when the fuel consistently takes the ``downhill'' step to waste~\cite{yanagida1985sliding,toleikis2020backstepping}.
Although many biophysical motors exist in the strong mechanochemical coupling regime, synthetic machines have only been built with loose coupling~\cite{borsley2024molecular,sangchai2023artificial}.
Understanding and optimizing these synthetic machines requires the development of design principles for stochastic motors with loose mechanochemical coupling.

A basic question is whether one can build a better motor by combining multiple inferior ones.
Suppose, for example, that a load-free motor drifts to the right with an average current \(j\).
Is it possible to physically couple two identical motors so that the dimer drifts to the right with a larger average current?
Due to the stochasticity, one motor will advance ahead of the other.
The leader will pull the laggard to catch up, but the laggard will pull back with an equal and opposite force in the other direction.
At first glance, the reciprocal forces between the leader and the laggard might suggest that the coupling must offer no benefit.
However, a force pulling a motor monomer in one direction can yield more motion than the same force pulling the monomer in the opposite direction.
Consequently, two motors mechanically pulling against each other with reciprocal forces can give rise to net motion in a favored direction.
Indeed, our manuscript gives an explicit illustration of how coupling two slow motors can give rise to a dimer with boosted current.

Our focus is on the type of catalysis-driven machine made by synthetic chemists.
The development of those synthetic molecular machines has progressed through several key milestones.
Early efforts focused on light-driven rotary motors, such as those pioneered by Feringa and co-workers~\cite{koumura1999light}, which achieved unidirectional rotation through photoisomerization of overcrowded alkenes.
Parallel work explored mechanically interlocked architectures like catenanes~\cite{dietrich1983nouvelle,hernandez2004reversible} and rotaxanes~\cite{anelli1991molecular,serreli2007molecular,crowley2009active}, which could undergo large-amplitude mechanical motions along well-defined pathways.
These structures were later mechanochemically coupled to fuel cycles, notably in the autonomous catenane motor~\cite{Wilson_2016} developed by the Leigh group, where catalysis of a fluorenylmethyloxycarbonyl chloride fuel drove the directional motion of the ring around a macrocycle. 
More recently, molecular motors capable of controlled rotation around a single covalent bond have been realized~\cite{borsley2022autonomous,wang2025transducing,collins2025}. 
These advances highlight the growing capacity to program complex dynamics into synthetic systems, and they raise the possibility of coupling multiple molecular machines to achieve collective behaviors~\cite{giuseppone2015gel1, giuseppone2017gel2, borsley2024gel3}.
Indeed, the underlying architectures\textemdash whether catenanes, rotaxanes, or single-bond rotors\textemdash are all structurally amenable to being linked together, suggesting a path toward supramolecular assemblies that exhibit cooperative functions.

Coupled synthetic machines are a particularly natural target given the variety of important functions performed by coupled subunits in biology.
The mechanical coupling between F$_\mathrm{O}$ and F$_\mathrm{1}$ is an important ingredient for energy transduction in ATP synthase~\cite{lathouwers2020nonequilibrium}, enzyme catalysis rates can be enhanced by combining multiple enzymes~\cite{agudo2021synchronization}, and multiple transport motors can be attached to a shared cargo to distribute the load~\cite{Huang_2025, Berger_2013, Rogers_2009, Gunther_2015}.
Even when moving without a load, coupled subunits are common in transport motors.
Kinesin, myosin, and dynein all have versions that can act as monomeric motors~\cite{okada1999processive,post2002myosin,sakakibara1999inner}, but most machines function as dimers, with two physically linked heads walking hand-over-hand~\cite{schliwa2003molecular}.
Linking the subunits admits the possibility that forces can be transmitted from one motor to another.
For example, the strain that builds up in a linker might help induce steps~\cite{Berger_2013}.
However, experiments with coupled motors that move in the absence of a load have suggested that the drift velocity is not appreciably affected by the coupling~\cite{Kunwar_2010,Gagliano_2010}.
Computational and theoretical models provide a setting for evaluating how generic those results might be.
In other words, under what conditions can coupling boost the current, and by how much?

Our approach uses two different classes of models.
Each model imagines unfurling the large ring of a catenane motor into a fixed linear track along which a shuttling ring moves.
A second shuttling ring moves along a parallel neighboring track, with the two shuttling rings physically coupled together.
That coupling could be simply modeled by an ideal spring (main text) or by an explicit polymer chain (supporting information).
In addition to the reciprocal mechanical forces between the shuttling rings, the models capture the reciprocal mechanochemical coupling needed to generate catalysis-driven motion.
Catalysis biases the statistics of the forces on a motor, and forces on a motor can also change the rate of catalysis.
For example, the ADP release step in myosin is known to be force-sensitive~\cite{Greenberg_2016}.

One way to capture reciprocal mechanochemical coupling is to use minimal models whose dynamics can be simulated based on essential physical interactions~\cite{ehrmann2025controlling,munoz2023computational,chatzittofi2025mechanistic}.
In our previous work, we developed a catalysis-driven motor model by leveraging explicit particle-based nonequilibrium molecular dynamics (MD)~\cite{Albaugh_2022, Albaugh_2023, Albaugh2023JCTC,penocchio2024power}.
We introduced chemostats that maintain a nonequilibrium concentration of a ``fuel'' whose catalytic decomposition into a ``waste'' can rectify the random motion of the motor. 
The particles comprising the system, fuel, and waste move via Langevin dynamics with forces derived from the gradient of a potential energy function.
The model is parameterized by pair potentials, so reaction rates for conversions between fuel and waste (including force-dependent reaction rates) emerge as a consequence.
Another route to capture the mechanochemical reciprocity is to model the one-dimensional (1D) diffusion of each shuttling ring on an energy landscape, with stochastic chemical events triggering jumps to different landscapes.
Such a jump-diffusion (JD) model requires reaction rates for the chemical jumps to vary as a function of the ring positions.

We deploy both classes of models in a complementary fashion to illustrate several main results.
First, both explicit MD and JD models clearly illustrate that coupling motors can boost current.
Second, the models clarify the conditions that are required to generate such a boost.
Mechanical coupling should reduce a transition-state barrier to make motor steps faster.
A typical outcome is faster steps but with a smaller bias favoring one direction.
The balance between those counteracting effects can enhance current, particularly if one can recover the lost bias by turning up the fueling. 
Third, our current boosts increase the current by only single-digit factors.
The JD results suggest that any prospects for boosting current by orders of magnitude require a change in design to dramatically increase the activity of the motors, likely by abandoning the core design element that a shuttling ring makes discrete rare hops between \emph{stable} binding sites.

\section*{Results}
\subsection*{A Current Boost in Explicit MD Simulations}
The first key result is an unambiguous confirmation that the laws of physics are consistent with a coupling-induced current boost.
In the Introduction, we highlighted that the mechanical coupling is reciprocal.
The leading and lagging motors pull on each other with equal and opposite forces, a balance that may appear to exclude a coupling-induced boost.
We therefore find it important to establish that a boost can be achieved in a simulation of explicit coupled motors.

By explicit, we mean that the dynamics of the motors are generated via forces derived as the gradient of a potential energy function.
This approach is to be contrasted with a kinetic rate model, which treats the rates for motor steps as predefined model parameters~\cite{penocchio2024power}.
In such a model, one can inadvertently adjust rates to enter parameter regimes that are thermodynamically inconsistent.
The explicit model excludes this concern.
One does not have independent control over all of the rates; they emerge as a consequence of conservative potentials and the resulting dynamics.
The technical details of both potentials and dynamics are described in Methods, and further numerical details are included in Supplementary Information (SI) Sec.\ 1.

\begin{figure*}[ht]
\centering
\includegraphics[width=1\textwidth]{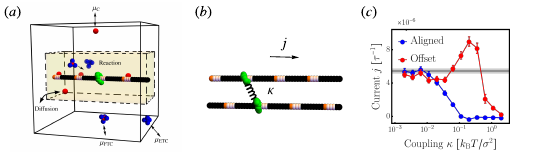}
\caption{
\textbf{Current boosts in coupled catalysis-driven linear motors.}
\textbf{(a)}
The system consists of a ring (green) diffusing along a track (black) with four equally spaced binding sites (orange) and adjacent catalytic sites (white).
The ring preferentially binds to the binding sites, while white catalytic sites catalyze the decomposition of full tetrahedral clusters (FTC) into empty tetrahedral clusters (ETC) and a central particle (C), which obstructs the ring's movement when attached to the catalytic site.
The track is infinitely extended using periodic boundary conditions. 
A nonequilibrium steady state is maintained by applying a high chemical potential for FTC (\(\mu_\mathrm{FTC}\)) and low chemical potential for waste products (\(\mu_\mathrm{ETC}\) and \(\mu_\mathrm{C}\)), driving net rightward motion of the ring.
\textbf{(b)} 
Two identical motors are arranged on parallel tracks in an offset configuration.
The centers of mass of the two rings are coupled by a harmonic potential with spring constant \(\kappa\). 
\textbf{(c)} When a single motor is operated under the nonequilibrium conditions, it generates an average shuttling-ring current \(j\) indicated by the gray line.
A boost above this uncoupled value is not possible when tracks are aligned but is possible for the offset configuration provided the coupling is strong enough but not too strong.
Means and error bars are collected from 100 independent simulations for each motor, all simulated with equal driving force $\Delta \mu = \mu_{\rm FTC} - \mu_{\rm ETC} - \mu_{\rm C}$ and plotted with reduced units defined in the main text.
Simulation methodology is discussed in Methods.
Numerical details about interaction strengths and driving forces are provided in the SI.
}
\label{fig:fig1}
\end{figure*}

To demonstrate the current boost with explicit MD simulations, we leveraged our previous work on a minimal particle-based model of an individual (uncoupled) motor~\cite{Albaugh_2022}.
The model captures motion of a linked ring of particles, the ``shuttling ring'', along a track.
In equilibrium, the shuttling ring diffuses along the track due to thermal fluctuations.
That motion can be rectified when mechanochemically coupled to a catalytic decomposition of a fuel, modeled again with a particle-based approach.
A tetrahedral cluster of particles, held together with harmonic bonds, can house a fifth particle inside the cluster.
This ``central particle'' (C) is sterically trapped until an appropriate fluctuation of the tetrahedral cluster enables an escape from a ``full tetrahedral cluster'' (FTC) to an ``empty tetrahedral cluster'' (ETC) and C.
Interactions between components of the motor's track and the tetrahedron catalyze that decomposition, thereby mechanochemically coupling the dynamics of shuttling-ring steps with fuel decomposition.
We have previously shown how our model can be simulated with Langevin dynamics and Monte Carlo chemostats that maintain a thermodynamic driving force that, on average, injects FTC and withdraws ETC and C~\cite{Albaugh_2022} (see also SI Sec.\ 2 for discussion of timescales and damping).
Although initially conceived as a model of a catenane motor with a shuttling ring moving along a fluctuating circular track, we have also studied a motor with a fixed linear track and periodic boundary conditions~\cite{Albaugh_2023}, as shown in Fig.~\ref{fig:fig1}(a).

Our initial numerical experiments place two parallel linear tracks next to each other.
Those tracks, each housing a single shuttling ring, are built from a chain of particles with diameter $\sigma$ and are separated by a distance $10 \sigma$.
The tracks do not exert forces on each other, but the shuttling rings do via a connecting spring with harmonic potential
\begin{equation}
  U_{\rm coupling} = \frac{1}{2} \kappa \Delta x^2
  \label{eq:harmonic_coupling}
\end{equation}
that aims to minimize the distance between the center of mass of each ring \(\Delta x\), with a spring constant \(\kappa\).
The tracks themselves have two periodically repeated features.
Some particles along the track (orange) attract the shuttling ring, providing binding sites where the shuttling ring dwells.
Other particles along the track (white) serve to catalyze the decomposition of the FTC and to bind the liberated C, which subsequently blocks the shuttling ring.
Within a track, the spacing of the binding sites and catalytic sites relative to each determines whether the shuttling ring current is positive, negative, or zero~\cite{Albaugh_2023}.
Here, we must also consider the importance of the relative positioning of the binding sites across the two tracks.
We configure each track's binding site and catalytic site spacing to prefer a rightward (positive) current under conditions with net FTC \(\to\) ETC + C decomposition.
Those tracks can be offset or aligned.
The offset option is the maximally out-of-phase configuration with binding sites positioned as illustrated in Fig.~\ref{fig:fig1}(b), while the aligned option is the in-phase configuration shown in Fig.~\ref{fig:figS1}(a).

Fig.~\ref{fig:fig1}(c) plots the average shuttling ring current \(j\) as a function of coupling strength \(\kappa\) for the aligned and offset configurations.
We work in reduced units in which particle diameter \(\sigma\), the particle masses \(m\), and the Boltzmann constant $k_{\rm B}$ are all set to unity.
Consequently, energies are naturally expressed in units of \(k_\mathrm{B}T\), the spring constant $\kappa$ in units of \(k_\mathrm{B} T / \sigma^2\), and times in units of \(\tau = \sqrt{m/(k_\mathrm{B}T)}\), where $T$ is the temperature.
For aligned tracks, Fig.~\ref{fig:fig1}(c) shows that the current decreases monotonically with increased coupling between the shuttling rings (further analysis in SI Sec.\ 3). 
By contrast, the offset tracks can achieve roughly 50\% more current when optimally coupled.
Notably, the current enhancement is not limited to harmonic spring coupling. 
We also performed numerical experiments with an explicit linker (a chain of particles) connecting the shuttling rings, described in SI Sec.\ 4.
The remainder of this paper focuses on the simpler harmonic coupling with the aim of dissecting the physical origin of the current boost.

\begin{figure*}[ht]
\centering
\includegraphics[width=0.9\textwidth]{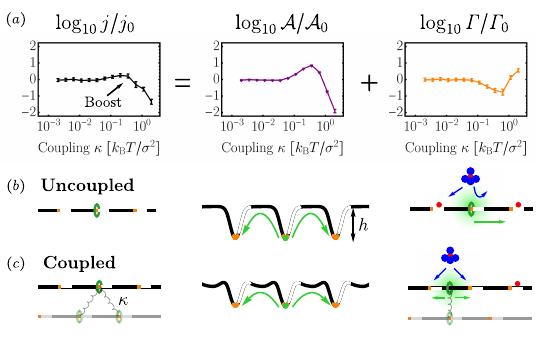}
\caption{
  \textbf{Current decomposition into activity and bias.}
  \textbf{(a)} For an offset motor configuration, current $j$ compared to its single-motor uncoupled value $j_0$ rises then falls as coupling between two motors is strengthened.
    This effect is productively rationalized by decomposing the coupled current into contributions from an activity ($\mathcal{A}$) and a bias ($\mathit\Gamma$) relative to the uncoupled values ($\mathcal{A}_0$ and $\mathit\Gamma_0$) then observing that the amplification of activity exceeds the loss of bias.
    \textbf{(b)} With no fuel and no coupling, the ring has symmetric jumps regulated by a \(h = 6.3\) \(k_{\rm B} T\) free energy barrier that the ring must jump over to get to a neighboring binding site.
    See SI Sec.\ 5 for free energy calculations of the MD model that generate the barrier-hopping cartoon.
  Adding fuel introduces a directional bias because the shuttling ring at a binding site sterically occludes (green shading) fuel from adding blocking groups at the nearby catalytic site.
  \textbf{(c)} Coupling two motors with offset tracks and an intermediate value of $\kappa$ yields shallower free-energy basins at the binding sites.
  More rapid escapes from those basins give an increase in activity, but they also increase the time the ring spends away from its binding sites, thereby decreasing the kinetic asymmetry that generates bias.} 
\label{fig:fig2}
\end{figure*}

\subsection*{Current, Bias, and Activity}

Having established that the numerical experiments can generate a coupling-induced boost in current, our attention shifts to explaining why the boost emerges and under what conditions it can be expected.
We can quantify the functioning of the motor in terms of the probability per unit time of hopping to neighboring binding sites, with a rate \(r_{\rm R}\) of moving to the right and \(r_{\rm L}\) of moving to the left.
Those rates can be extracted from the numerical experiments by counting the hops realized over a simulation time (see Methods).
We find that it is productive to factorize \(j = r_{\rm R} - r_{\rm L}\) in a seemingly tautological manner:
\begin{equation}
  j = (r_{\rm R} + r_{\rm L}) \left(\frac{r_{\rm R} - r_{\rm L}}{r_{\rm R} + r_{\rm L}}\right),
  \label{eq:decomposition}
\end{equation}
where the first term $\mathcal{A} \equiv r_{\rm R} + r_{\rm L}$ is a time-reversal-symmetric \emph{activity} that sets a timescale for hops and the second term $\mathit\Gamma \equiv (r_\mathrm{R} - r_\mathrm{L})/(r_\mathrm{R} + r_\mathrm{L})$ is a time-reversal-antisymmetric \emph{bias} ranging from -1 (pure leftward) to +1 (pure rightward). 
An advantage to such an activity-bias decomposition is that we can separately reason about the impact of coupling on the two terms.

The activity \(\mathcal{A}\) is quantified in terms of rates extracted from nonequilibrium dynamics.
Despite this complexity, the response of activity to coupling can be anticipated by
an analysis of unfueled diffusion on an equilibrium free-energy landscape.
The free energy minima correspond to binding sites for the ring, and transitions between those minima require climbing over a free energy barrier.
Arrhenius barrier-crossing rates are exponentially sensitive to the free-energy difference between the minimum and the barrier, introducing a way for small changes in the landscape to induce big changes to the activity.
If coupling two motors can destabilize the binding site, even slightly, the sizable jump in activity could be enough to boost the current.
The cartoon in Fig.~\ref{fig:fig2} illustrates how coupling motors with offset tracks leads to shallower minima, but there is a limit to the effect, shown explicitly in Sec.\ 5 of the SI.
If the spring is too strong, the intermediate state (with shuttling ring between two binding sites) becomes metastable, in which case transit through the state slows.
Of course, when tracks are aligned, all values of $\kappa$ lead to deeper wells (see SI Sec.\ 3), which is why Fig.~\ref{fig:fig1} lacks a current boost for aligned tracks.

The equilibrium picture helps clarify why coupling can boost activity, but it does not clearly indicate whether it might degrade bias by a similar amount.
In fact, our numerical experiments show that changes in activity and bias often counteract each other, as illustrated in Fig.~\ref{fig:fig2}.
Reasoning about the bias requires one to consider the kinetic asymmetry~\cite{astumian1989,astumian2019,penocchio2024multicycle} in the different rates for reactions to be catalyzed by the different motor configurations.
Most importantly, C particles can be extracted from bulky FTC and deposited at catalytic sites, with a rate that dramatically decreases if the shuttling ring is close enough to the catalytic site to sterically block access.
The feedback between the position of the ring and the rate of adding blocking groups induces biased motion through the information ratchet mechanism~\cite{astumian2019,sangchai2023artificial}.
That information ratchet reasoning leans heavily upon an analysis of the differences in catalytic rates when the ring is bound to or unbound from a binding site.
If the bound configuration is significantly disfavored, the kinetic asymmetry is lessened because the ring is rarely close enough to a catalytic site to create asymmetry in the rate at which blocking groups are deposited.
Because the offset configuration's spring pulls both shuttling rings away from their binding sites, the increase in activity is therefore accompanied by a degraded bias.

Although coupling impacts activity and bias in counteracting directions, those effects need not be balanced.
That is, the benefit from increasing activity can exceed the loss from reduced bias.
The balance, or imbalance, as it is, is especially clear by viewing activity and bias on a logarithmic scale and comparing with the uncoupled values (subscript 0):
\begin{equation}
  \log \frac{j}{j_0} = \log \frac{\mathcal{A}}{\mathcal{A}_0} + \log \frac{\mathit\Gamma}{\mathit\Gamma_0}.
\end{equation}
When the activity boost exceeds the bias degradation, coupling will speed up the collective relative to the individual.

\subsection*{Offset Tracks are Necessary but Not Sufficient to Boost Current}

For coupling to enhance activity, a spring must coax a shuttling ring to pull away from its otherwise stable binding site.
Coupled shuttling rings on two aligned tracks can sit simultaneously on their binding sites, but offset tracks introduce frustration.
For those offset tracks, the spring stabilizes the transition state with one ring comfortably bound to a binding site, and the other ring sitting between binding sites.
That stabilization is essentially baked into the geometry, yet the offset geometry alone is not sufficient to guarantee a current boost.
The depth of the binding site free energy minima also matters.

\begin{figure}[t!]
\centering
\includegraphics[width=0.5\textwidth]{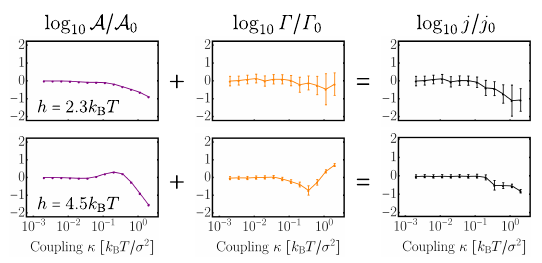}
\caption{ 
\textbf{Current is not boosted when the shuttling ring is weakly attracted to binding sites.}
  When the uncoupled motor had a \(6.3 k_{\rm B} T\) deep attractive well holding the shuttling ring to each binding site, Fig.~\ref{fig:fig2} showed that the coupling-induced boost in activity (\(\mathcal A/\mathcal A_0\)) could be greater than the corresponding loss in bias (\(\mathit \Gamma/\mathit \Gamma _0\)).
  When the binding sites are less attractive, a current boost was not observed.
  Very weak binding of shuttling ring to the binding sites ($h = 2.3 k_{\rm B} T$) allows for such rapid hops that coupling cannot increase activity, much less current.
  Being coupled to another ring can boost the activity when $h = 4.5 k_{\rm B} T$, but this boost is not as great as the loss in bias.
  For these rotaxanes to get a boost, it is important that the well depth is sufficiently deep that the ring spends an overwhelming fraction of the time at a binding site.
}
\label{fig:fig3}
\end{figure}

Consider, for example, the limit that the free-energy minima are shallow relative to thermal energy $k_{\rm B} T$.
Then, the activity does not have a simple Arrhenius picture, with the coupling modulating the depth of the basins.
Consequently, the coupling of offset motors may not even increase the activity, an effect that we observe in the top panel of Fig.~\ref{fig:fig3} when the well depths are $h = 2.3 k_{\rm B} T$.
As the binding sites are made more attractive, the shuttling-ring kinetics begins to adopt discrete Arrhenius-like hops.
At that point, coupling stabilizes the transition to allow a boost in activity.
Even when wells are deep enough to generate an activity boost, there is no guaranty that enhanced activity exceeds the drop in bias.
In fact, Fig.~\ref{fig:fig3} shows that loss of bias can exceed gain of activity when $h = 4.5 k_{\rm B} T$.

The current boost described earlier in the paper emerged when the orange and green particles were so attractive that $h = 6.3 k_{\rm B} T$.
Coupling the motors draws the shuttling rings away from their binding sites, increasing the activity as before.
However, with deeper well depth, the shuttling rings spend much more time in the bound configurations, where the kinetic asymmetry of the information ratchet mechanism generates a strong directional bias.
It is a subtle balance between counteracting effects, but Fig.~\ref{fig:fig2} shows that the enhancement of activity is the dominant effect for deep wells.

\begin{figure*}[t!]
\centering
\includegraphics[width=0.8\textwidth]{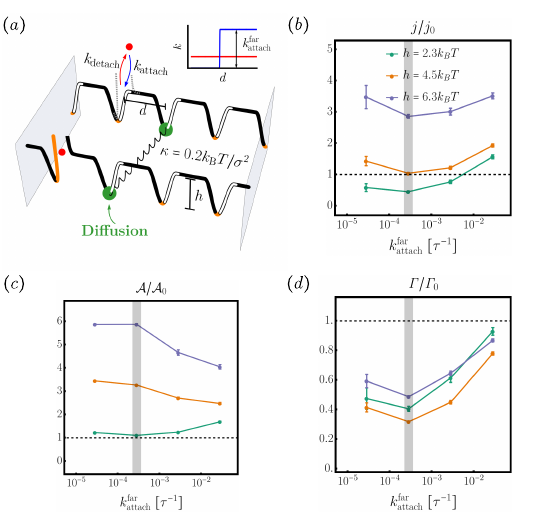}
\caption{
\textbf{One-dimensional jump-diffusion model to study the effect of fuel concentration.}
\textbf{(a)} 
 The ring (green circle) diffuses along a one-dimensional potential fitted to MD simulations with parameter \(h\) setting the barrier height. 
 Motors on parallel offset tracks are coupled by a harmonic potential. 
Blocking groups (red circle) are added or removed through a jump process with attachment and detachment rates, \(k_\mathrm{attach}\) (red) and \(k_\mathrm{detach}\) (blue).
\(k_\mathrm{attach}\) depends on the distance $d$ between the ring and the catalytic site (white), modeled by a function that switches from zero at short distances to \(k_\mathrm{attach}^\mathrm{far}\) at long distances.
\textbf{(b)} The coupling-induced current amplification \(j/j_0\) at coupling strength \(\kappa = 0.2k_\mathrm{B}T / \sigma^2\) as a function of \(k_\mathrm{attach}^\mathrm{far}\) with the $h$ parameter set to match the MD barrier heights of Figs.~\ref{fig:fig2} and~\ref{fig:fig3}.
The gray line highlights the corresponding \(k_\mathrm{attach}^\mathrm{far}\) simulated in the MD simulation.
Increasing $k_{\rm attach}^{\rm far}$ can generate a current enhancement even for small well depths.
\textbf{(c)} Deeper wells consistently yield larger activity boosts across \(k_\mathrm{attach}^\mathrm{far}\).
\textbf{(d)} The current boost emerges because increasing \(k_\mathrm{attach}^\mathrm{far}\) restores the bias towards its uncoupled value while largely preserving the coupling-boost activity boost.
Means and error bars are collected from 100 independent simulations for each motor. 
}
\label{fig:fig4}
\end{figure*}

\subsection*{Recovering a Current Boost By Increasing the Fueling Rate}

So far, we have considered tuning three structural features of the motors: (1) the alignment of tracks, (2) the strength of coupling between two linked shuttling rings, and (3) the strength of attraction that holds a shuttling ring to a binding site.
In the numerical experiments described so far, we have kept the properties of the environment fixed\textemdash the temperature and, even more importantly, the chemical potential drop from the consumption of an FTC and the generation of an ETC and C.
One possible way to adjust the driving force is to alter the concentration of FTC, and this additional control gives the flexibility to engineer conditions for boosting current in two steps.
First, the coupling can be adjusted to boost activity.
Second, the fueling conditions can be adjusted to increase bias.
Essential to this two-step strategy is the assumption that increasing the fuel concentration restores bias without significantly degrading activity.

Validating that assumption is not particularly straightforward with the MD model.
At higher fuel concentrations, the computational cost increases, making systematic investigations of concentrations across several orders of magnitude challenging.
This difficulty motivates the development of a complementary jump-diffusion (JD) model to probe the dependence on the FTC concentration.
As in the MD model, the JD model incorporates continuous diffusion of the ring along a periodic track due to forces from a potential energy that depends on the blocked or unblocked status of the catalytic sites.
The difference is that toggling between blocked and unblocked sites is handled at the level of a kinetic model.
There are rates of jumping from blocked to unblocked, and back.
The mechanochemical coupling between these jumps and the ring diffusion enters because the jump rates are modeled as being dependent on the ring's position. 
In SI Sec.\ 6, we explicitly demonstrate that appropriate parameterization of this JD model recapitulates the features of the MD model. 

In the JD model, each ring is shrunk to a point that diffuses along a 1D line with a potential informed by the free energy for the center of mass of the MD shuttling ring.
Along with forces generated from the gradients of the 1D potentials, the diffusing particles pull on each other with a harmonic coupling force (see also Methods).
In the absence of blocking groups, this potential resembles the cartoon in the middle of Fig.~\ref{fig:fig2}(b) with \(h\) setting the height of the barrier that regulates the escape of the shuttling ring from a binding site.
When a catalytic site is blocked, the potential spikes, mimicking the steric repulsion between the shuttling ring and C particles at catalytic sites that gate the ring's diffusion.
From the ring's perspective, the potential around a binding site thus jumps between a blocked and unblocked energy landscape, depicted in Fig.~\ref{fig:fig4}(a).
We define \(k_{\rm attach}\) and \(k_{\rm detach}\) as the attachment and detachment rates for the blocking groups.
Consistent with MD results of Fig.~\ref{fig:figS5}, \(k_{\rm detach}\) is modeled as being independent of the distance between the shuttling ring and the catalytic site (where detachment occurs).
By contrast, \(k_{\rm attach}\) is a distance-dependent rate because steric repulsions between the shuttling ring and the FTC affect the accessibility of the catalytic site.
We approximate the rate of attachment to be given by a switching function that is essentially zero if the shuttling ring is too close to the catalytic site.
Beyond a critical distance (essentially the size of FTC in the MD model), our JD model assumes that FTC can access the catalytic site and the blocking rate no longer depends on the distance $d$.
That large-$d$ limiting rate, $k_{\rm attach}^{\rm far}$, can be thought of as a measure of the catalytic decomposition rate, since it sets the timescale for adding blocking groups via the catalytic decomposition of FTC.
In an MD model, adjusting that rate requires either an increase in FTC concentration or a set of pair potentials that are more catalytically active.
For the JD model, we think of $k_{\rm attach}^{\rm far}$ as a parameter that sets the FTC concentration.

Using the JD model, we focused on the offset track configuration with an intermediate coupling strength \(\kappa = 0.2k_\mathrm{B}T/\sigma^2\), the conditions that the MD simulations had shown to be capable of generating the largest boost.
JD simulations were performed with the barrier height parameter $h = 2.3, 4.5,$ and $6.3 k_{\rm B} T$ corresponding to the MD conditions, all while varying $k_{\rm attach}^{\rm far}$.
Fig.~\ref{fig:fig4}(b) shows how $k_{\rm attach}^{\rm far}$ impacts coupling-induced current amplification $j / j_0$.
As in the MD simulations, deeper wells enhance current boosts, yet increasing fuel concentration can also elicit boosts in shallow-well motors.
Fig.~\ref{fig:fig4}(c) confirms that these current enhancements again arise from increased activity, with a greater activity boost the deeper the well.
Those activity boosts are accompanied by a degraded bias with $\mathit \Gamma / \mathit \Gamma_0 < 1$, but in the limit of large $k_\mathrm{attach}^\mathrm{far}$, the motor tends toward its original uncoupled bias.
Together, these results establish a general strategy: first, tune the coupling to increase activity, and then restore bias by increasing the fuel concentration.

\section*{Discussion}
Rotaxane and catenane-based synthetic machines have made great strides in designing architectures that convert fuel decomposition into directional bias.
However, generating that bias alone does not make a great motor.
It is necessary for the motor to be sufficiently active that the bias is translated into a high current.
For example, the initial synthetic autonomous catenane motor succeeds in generating biased motion, but only executes around 2 net cycles per day~\cite{Wilson_2016}.
Perhaps the most natural way to try to improve the motor is to add more fuel, increasing the thermodynamic driving force.
For a small driving force, one generically expects a linear regime, wherein the current grows proportionally to the thermodynamic driving force.
Unfortunately, increasing the driving force does not generically produce a high-current motor because the response to fuel saturates~\cite{yasuda1998f1} (see also SI Sec.\ 7).
We need additional strategies to make motors generate more current when simply adding more fuel will not work.

Our numerical investigations support a two-step approach to speeding up motors by coupling multiple units.
The first step is to devise a strategy to speed up the activity by lowering a transition state.
Here, this strategy meant using offset tracks and linking motors with a harmonic mechanical coupling that is strong enough but not too strong, akin to the intermediate coupling regime previously found to optimize energy transduction between strongly coupled subunits~\cite{lathouwers2020nonequilibrium}.
The coupling enhances activity but decreases bias, so that the new motor complex is no longer saturated with regard to increasing fuel concentration.
The second step is to add more fuel to recover lost bias while maintaining enhanced activity.

Our models show that current could be boosted by a factor of 3 to 4 under the most favorable conditions we explored, deep binding wells and high fuel concentration.
Our findings broaden the design elements that synthetic chemists should consider when building molecular machines.
The first instinct is often to optimize individual motors by modifying their chemical structures, an approach that has generated biased motion but limited current.
Coupling might provide a way to turn those precise but slow motors into faster machines.
At the same time, our results highlight that amplifying current by orders of magnitude would require new strategies, likely requiring designs to move away from architectures with rare hops between stable binding sites.

\section*{Methods}
\subsection*{Particle-based Model}
\subsubsection*{Dynamics}
The particle-based model builds on previously described work~\cite{Albaugh_2022,Albaugh_2023,penocchio2024power}.
Briefly, the diffusive dynamics of the particles in the motor system was simulated with underdamped Langevin dynamics.
At each timestep, particles have a momentum and experience forces due to other particles as well as receiving a random force, drawn from a Gaussian, to model the effects of an implicit solvent.
For each particle \(i\), the position $\mathbf{r}_i$ and the momentum $\mathbf{p}_i$ evolve according to
\begin{equation}
\begin{aligned}
    \dot{\mathbf{r}}_i(t) =& \frac{\mathbf{p}_i(t)}{m_i}, \\
    \dot{\mathbf{p}}_i(t) =& -\nabla_i U(\mathbf{r}) -\frac{\gamma}{m_i}\mathbf{p}_i(t)+\xi_i(t).
\end{aligned}
\label{eq:Langevin}
\end{equation}
Here (\(\,\dot{}\,\)) is the time derivative, \(m_i\) is the mass, \(\nabla_iU\) is the gradient of potential energy of the whole system with respect to the position of particle \(i\), \(\gamma\) is the friction coefficient, and \(\xi_i\) is the white noise independently drawn from the Gaussian distribution with mean zero and variance \(2 \gamma k_{\rm B} T\).
As described in SI Sec.\ 1, typical simulations were performed with \(T = 0.5\), \(m = 1\), and \(\gamma =0.5\) for \(2 \times 10^8\) timesteps of size \(\Delta t = 5 \times 10^{-3}\), all expressed in the reduced units with \(\sigma = m = k_{\rm B} = 1\).
For technical reasons, it is beneficial to numerically integrate the underdamped Langevin dynamics with a BAOAB integrator~\cite{leimkuhler2013rational,fass2018quantifying}.
Despite the use of the underdamped integrator, it is important to note that for the chosen parameters the motor motion is effectively overdamped, as clarified in the SI Sec.\ 2.

In parallel with the Langevin dynamics, the constant concentrations of FTC, ETC, and C were maintained through grand canonical Monte Carlo (GCMC) chemostats that inserted and deleted the species in a cubic ``outer box'' with side length \(34 \sigma\).
Those species rapidly diffuse through a cubic semi-permeable ``inner box'' with side length $30 \sigma$, within which the motor is confined.
The concentrations of FTC, ETC, and C in the inner box are thus controlled without having insertion or deletion moves in the immediate vicinity of the motor.
A GCMC move was attempted every 100 Langevin timesteps, a frequency that is fast enough that the concentrations in the inner box always reflect the chemical potentials applied to the outer box~\cite{Albaugh_2022}.

\subsubsection*{Interactions}
The deterministic forces acting on the particles are derived from the gradient of a potential energy function.
All particles interact with Lennard-Jones (LJ) pair potentials to mimic long-range attractive van der Waals forces and short-range repulsive exclusion forces:
\begin{equation}
\begin{aligned}
    U_\mathrm{LJ}(\mathbf{r}_{ij}) = 4 \epsilon_{R,ij} \left( \frac{\sigma_{ij}}{|\mathbf{r}_{ij}|}\right)^{12} - 4 \epsilon_{A,ij} \left( \frac{\sigma_{ij}}{|\mathbf{r}_{ij}|}\right)^{6}
\end{aligned}
\label{eq:LJ}
\end{equation}
where \(\mathbf{r}_{ij}\) is the distance vector between particles \(i\) and \(j\), \(\epsilon_{R,ij}\) regulates short-range repulsions, \(\epsilon_{A,ij}\) regulates long-range attractions, and \(\sigma_{ij}\) is the sum of the radii of particles \(i\) and \(j\).
The LJ pair potentials are supplemented by two additional types of pairwise interactions.
The particles that form the tetrahedral cage in FTC and ETC (blue) are held together by harmonic bonds.
Bonds, in this case finitely extensible nonlinear elastic (FENE) bonds, are also added to hold together neighboring beads along the track, within the shuttling ring, and in the explicit linker described in SI Sec.\ 4.
In addition to the pair potentials, a many-body angular potential is included to impose a roughly circular shape on the shuttling ring.
Details of all terms of the potential energy are discussed further in SI Sec.\ 1, including values of the parameters defining those energy functions.

A significant part of this work required that we tune the depth of the free-energy wells that hold the shuttling ring at the binding sites.
We achieved this by setting the attractive LJ parameter $\epsilon_{\rm A}$ between green and orange particles to 0.8, 1.4, and 1.8\(k_{\rm B} T\), producing the low (\(2.3 k_{\rm B} T\)), medium (\(4.5 k_{\rm B} T\)), and high (\(6.3 k_{\rm B} T\)) well depths.
Notice that we actually tune $\epsilon_{\rm A}$ and simply measure the resulting well depth $h$ from a single-track, single-shuttling-ring free energy calculation (black curve in Fig.\ S3).

\subsubsection*{Extraction of Hopping Rates}
The decomposition of the shuttling ring current into activity and bias requires extracting the hopping rates to neighboring binding sites, \(r_{\rm R}\) and \(r_{\rm L}\), from MD simulations.
To do so, we must coarse-grain to characterize frames of the MD trajectory as having the shuttling ring bound to or unbound from a binding site.
During a transition from bound to unbound, many transient recrossings can occur, obscuring the statistics of hopping rates.
To avoid over-counting those barrier recrossings, we adopted a so-called core set approach~\cite{schutte2011markov}.
Starting with the ring at a binding site, we register a hop once the center of mass of the ring first reaches one of the neighboring binding sites.
The ring then wiggles around that binding site, and we do not count small wiggles as additional ``hops''.
We only register another hop once the ring has again reached a different binding site.

There is a little bit of finesse in defining what it means for a ring to have reached a new binding site, and the choice impacts a decomposition into activity and bias.
Imagine that we registered hops once the center of mass passed through a region around the binding site with diameter $\sigma$.
Alternatively, we could consider a ``core set'' which is twice as large with diameter $2 \sigma$.
These two choices will provide different accounting for the hops observed by certain rare trajectories.
Suppose, for example, that the ring took an excursion from one binding site to the right, and its center of mass reached the particle right before the next binding site, but then came back.
By the 2-$\sigma$ accounting system, that excursion would register as a right hop followed by a left hop, contributing two hops to a count of activity and bringing the bias closer to a 50/50 split.
With the 1-$\sigma$ accounting system, that excursion would not register at all.
Notice, however, that the current is not sensitive to the accounting system; the current only cares about net motion and not about the manner in which it is partitioned into hops.

Although activity and bias are more sensitive than current to the size of the core set, they are not \emph{particularly} sensitive provided that the core sets are chosen in the right Goldilocks regime\textemdash not too big and not too small.
If the core sets are too large, one overcounts hops due to recrossings, recording wiggles far away from any binding site as if the system were hopping between two binding sites.
If the core sets are too small, one undercounts, particularly when the spring strains are such that the minimum of the free energy is not exactly at the binding site.
Provided the wells are sufficiently deep that there is a clear separation of timescales, the activity-bias decomposition is quite robust and a broad range of core set sizes (anything on the order of the width of the well) is fine.
For the activity-bias decompositions of MD data in Figs.~\ref{fig:fig2} and~\ref{fig:fig3}, we used a core set of size $\sigma$.
The decomposition of JD model data needed to also handle higher fueling rates.
In SI Sec.\ 8, we demonstrated that slightly larger core sets were more robust for the JD model across the range of fueling rates, so $3 \sigma$ core sets were used to generate Fig.~\ref{fig:fig4}.

\subsection*{Jump-Diffusion Model}
\subsubsection*{Ring's Dynamics}
The JD model replaces the explicit track and shuttling ring with an implicit one.
The ``ring'' becomes a single particle that diffuses along on a 1D periodic track potential with underdamped Langevin dynamics.
\begin{equation}
\begin{aligned}
    \dot{x}(t) =& \frac{p(t)}{M}, \\
    \dot{p}(t) =& -\frac{\partial U}{\partial x} -\frac{\gamma}{M}p(t)+\xi(t).
\end{aligned}
\label{eq:1DLangevin}
\end{equation}
As before, the dynamics involves a mass, damping coefficient, and random force.
Now, the 12-particle shuttling ring has been concentrated to the point-particle with $M = 12$.
The temperature $T = 0.5$ matches that of the explicit particle simulations, and the damping $\gamma = 6$ was chosen to give an equilibrium activity $\mathcal{A}$ roughly in line with the activity of the unfueled single-track explicit particle simulations (see SI Sec.\ 6).

The energy $U$ is dominated by the bare-track potential $U_0$ for the single ring's motion without any fuel or blocking groups.
That $U_0$ was parametrized to resemble the free-energy as a function of track position for the unfueled, single-track particle simulation.
There are wells at the binding sites that are patched together with flat diffusive barriers in a manner that is smoothly differentiable.
Specifically,
\begin{equation}
U_0(x) =
\begin{cases}
\frac{h}{2} & \text{if } \frac{w}{2}<\tilde{x}_i  \leq l-\frac{w}{2}, \\
\frac{h}{2} \sin [\frac{2\pi}{l} (\tilde{x}_i-\frac{l}{4})]  & \text{if } -\frac{w}{2}<\tilde{x}_i  \leq \frac{w}{2},
\end{cases}
\label{eq:1Dpotential}
\end{equation}
where \(\tilde{x} = [(x_i+w/2-\phi)\mod l ]- w/2\) is a transformed coordinate that imposes periodicity with phase $\phi$.
Setting $w = 6\sigma$ and $l = 12\sigma$, this $U_0$ adopts the form plotted at the top of Fig.~\ref{fig:fig4}(a), with wells 6 units wide separated by flat barriers 6 units wide.
The barrier height parameter $h$ was chosen to match barrier heights of the MD simulations, as discussed in SI Sec.\ 5.

Upon introducing a second track, the phase shift \(\phi\) is necessary to allow the tracks to be offset relative to each other.
The maximum phase-shifted ``offset'' configuration has $\phi = 0$ for track 1 and $\phi = l/2$ for track 2.
We couple rings on the two tracks with the same harmonic potential used for the explicit particle simulations, $U_{\rm coupling}$ in Eq.~\eqref{eq:harmonic_coupling}.
Although the ring moves in 1D, the distance \(\Delta x\) in $U_{\rm coupling}$ is a Euclidean distance in 2D assuming that the tracks are separated by a distance of $10 \sigma$.

Up to this point, we have described only a model for two coupled diffusive rings.
We also need the rings to feel a potential that jumps between different options when the catalytic sites are blocked or unblocked.
We explicitly track the blocked and unblocked state of the catalytic sites next to four wells (other wells are merely periodic replicas).
Every blocked site is modeled as if a blocking group is bound to the track at a position $\tilde{x} = 2 \sigma$, and that blocking particle imposes a repulsive potential felt by the ring.
For the explicit particle simulation, the interactions between a blocking C particle and the shuttling ring had no attractive part ($\epsilon_{\rm A} = 0$ to turn the LJ potential into a purely repulsive $r^{-12}$ potential).
We apply the same potential for the impact of the blocking group on the JD model, but we compute the distance for the repulsive potential simply as the 1D length along the track between the ring's position and the blocking group.
Because the ring's location has been concentrated down to a point lying \emph{on} the track, the JD model measures the blocking group and ring to be closer to each other than the corresponding particle simulation.
To roughly compensate, we reduce $\epsilon_{\rm R}$ from the particle simulation value of $2\times 10^5 k_\mathrm{B}T$ to $2\times 10^3  k_\mathrm{B}T$.

Given a pattern of blocked and unblocked catalytic sites on each track, we have thus described a potential energy, the derivative of which captures the forces felt by both rings.
This potential sums $U_0$ for each ring, $U_{\rm coupling}$ between the rings, and the repulsive $\epsilon_{\rm R} r^{-12}$ potential from each blocked catalytic site.
Ring positions and momenta are propagated again using BAOAB integrator with timesteps of size $\Delta t = 5 \times 10^{-3}$.
As in the particle-based model, the damping of the underdamped Langevin dynamics was sufficiently great that the ring dynamics effectively lacks inertial effects.

\subsubsection*{Kinetics of Chemical Reactions}
The final step is to describe how the position of the rings feeds back to impact the dynamics of the blocking groups.
After 100 steps of the ring dynamics, Monte Carlo moves were employed to randomly alter the blocking groups.
Recall that each track has catalytic sites that can be blocked or unblocked.
The reactions occur as instantaneous jumps in which each catalytic site can switch between its blocked or unblocked states.
Each site switches from blocked to unblocked with probability $100 k_{\rm detach}(d) \Delta t$, where $100 \Delta t$ is the time between attempted jumps and $k_{\rm detach}(d)$ is the detachment rate of Fig.~\ref{fig:fig4}(a) that depends on the distance between the ring and the catalytic site, $d$.
Unblocked sites similarly switch to blocked sites with probability $100 k_{\rm attach}(d) \Delta t$, a sufficient Monte Carlo frequency as shown in SI Sec.\ 8.
Practically, this flipping is achieved by measuring the distance between the ring and catalytic site, computing the corresponding flipping probability for each site, then drawing a random number on $\left[0, 1\right)$ to compare with the probability to flip the site.

A consequence of the flipped state of the catalytic sites is that the energy of the system can jump because there is a change in the potential between the blocking group and the ring.
The Langevin integrator naturally samples equilibrium dynamics at temperature $T$ (in the small $\Delta t$ limit), and one may worry that the jump processes will disturb that equilibrium unless $k_{\rm attach}(d)$ and $k_{\rm detach}(d)$ are carefully constructed to preserve detailed balance.
Indeed, if the goal were to sample equilibrium, it would be necessary to relate ratios of attachment and detachment rates to the Boltzmann weight associated with the binding of a blocking group to a catalytic site.
While one could imagine an arbitrary distance-dependent attachment rate, equilibrium thermodynamics would constrain the form of the distance-dependent detachment rate and vice versa.
We emphasize that our simulations need not (and do not) satisfy such an equilibrium relationship.
We explicitly intend for the choice of $k_{\rm attach}(d)$ and $k_{\rm detach}(d)$ to push the system out of equilibrium, an effect that emerges physically because the rates emerge out of a system that is consuming fuel (FTC $\to$ ETC + C).
The typical addition of a blocking group consumes a fuel molecule while the typical detachment of the blocking group does not regenerate the fuel.

The particle-based simulations make the net fuel consumption clear.
Those simulations show that blocking groups can become associated with catalytic sites via two distinct mechanisms: (1) an FTC can associate to the catalytic site, deposit a C as a blocking group, and release an ETC or (2) a free C can simply bind to a blocking group.
Both mechanisms are present along with their reverse process, but both mechanisms are not equally likely \emph{under the nonequilibrium fueling conditions}.
We are interested in the kinetics when FTC is held at a high concentration and ETC and C are held at low concentrations.
In that case, with overwhelming probability blocking groups add via mechanism (1) and remove via the microscopic reversal of (2).
Thermodynamic considerations restrict the ratio of rates for adding via (1) to removing via (1).
They also restrict the ratio of rates for adding via (2) to removing via (2).
Our $k_{\rm attach}(d)$ is a phenomenological form for a distance-dependent rate for addition via (1) while $k_{\rm detach}(d)$ is a phenomenological form for a distance-dependent rate for removal via (2), rates that are not constrained by thermodynamics.

Attachment occurs with a rate which is essentially zero if the ring's steric repulsion prevents an FTC from reaching a catalytic site.
Provided the FTC can approach, the attachment rate does not strongly depend on the distance to the ring.
We capture both effects by modeling $k_{\rm attach}(d)$ with a switching function to smoothly pass from 0 to $k_{\rm attach}^{\rm far}$ around $d \approx a$, with $a = 4.6$ being a length scale that quantifies the steric size of FTC.
The sharpness of the switch is regulated by a second parameter, $b = 0.001$ in
\begin{equation}
  k_{\rm attach}(d) = \frac{k_{\rm attach}^{\rm far}}{1 + e^{(a-d)/b}}.
\end{equation}
  The phenomenological rate for removal is modeled as a constant given that the detachment of C in MD simulations is barely affected by the ring's position, 
\begin{equation}
  k_{\rm detach}(d) = 1.5\times 10^{-4}.
\end{equation}
As discussed in SI Sec.\ 6, parameters of the rate model were chosen to fit the particle-based MD simulations.
\section*{Data Availability}
The datasets generated and analyzed during the current study are available in the Zenodo repository at https://doi.org/10.5281/zenodo.18237508.

\section*{Code Availability}
The source code and analysis scripts used in this study are available in a Zenodo repository under the DOI https://doi.org/10.5281/zenodo.18237508.
\subsection*{Acknowledgments}
We acknowledge support from the Gordon and Betty Moore Foundation under Grant Number GBMF10790.
This material is based upon work supported by the U.S. Department of Energy, Office of Science, Office of Basic Energy Sciences program under Award Number DE-SC0026333.
This research was supported in part by grants from the NSF (DMS-2235451) and Simons Foundation (MPS-NITMB-00005320) to the NSF-Simons National Institute for Theory and Mathematics in Biology (NITMB).

\bibliography{ref_ref}

\renewcommand{\thefigure}{S\arabic{figure}}
\setcounter{figure}{0}
\renewcommand{\thetable}{S\arabic{table}}
\setcounter{table}{0}
\renewcommand{\theequation}{S\arabic{equation}}
\setcounter{equation}{0}

\onecolumngrid

\newpage

\section*{Supporting Information}
\section{MD Simulation Details}
The details of MD models and simulation methods are essentially identical to those described previously~\cite{Albaugh_2023}. 
Here, we provide enough information to reproduce the MD simulations, but we also refer interested readers to Ref.~\cite{Albaugh_2022} for the development of the methodology and the SI of Ref.~\cite{Albaugh_2023} for more numerical details.
\subsection{Dynamics}
We numerically integrated Eq.~4 in the main text with the BAOAB integrator \cite{leimkuhler2013rational,fass2018quantifying},
\begin{equation}
\begin{aligned}
    \mathbf{p}_i^{j+\frac{1}{2}} =&\mathbf{p}_i^j + \frac{\Delta t}{2} \mathbf{f}_i^j\\
    \mathbf{r}_i^{j+\frac{1}{2}} =&\mathbf{r}_i^j + \frac{\Delta t}{2 m_i}\mathbf{p}_i^{j+\frac{1}{2}}\\
    \mathbf{p}_i^{j+\frac{1}{2}*} =& e^{-\frac{\gamma \Delta t}{m_i}} \mathbf{p}_i^{j+\frac{1}{2}} + \sqrt{1-e^{-\frac{2\gamma\Delta t}{m_i}}}\mathbf{\eta}_i^{j}\\
    \mathbf{r}_i^{j+1} =& \mathbf{r}_i^{j+\frac{1}{2}}+\frac{\Delta t}{2 m_i}\mathbf{p}_i^{j+\frac{1}{2}*}\\
    \mathbf{p}_i^{j+1} =& \mathbf{p}_i^{j+\frac{1}{2}*} +\frac{\Delta t}{2}\mathbf{f}_i^{j+1},
\end{aligned}
\label{eq:integrator}
\end{equation}
where \(j\) denotes the \(j^\mathrm{th}\) time step, \(\Delta t\) is the step size, \(\mathbf{f}_i^j = -\nabla_i U (\mathbf{r}_i^j)\) is the force on particle \(i\), and \(\eta_i^j\) is a random vector whose components are Gaussian variables with zero mean and unit variance.
Currents, activities, and biases were computed by averaging over 100 trajectories, each with \(2\times 10^8\) steps generated with \(\Delta t= 5\times 10^{-3}\), \(T = 0.5\), and \(k_\mathrm{B} = 1\).
\subsection{Potential}
In addition to the LJ potential described in the main text (Eq.~5), particles also experience the following potential interactions to form functional moieties (tetrahedral shell of the fuel, shuttling ring and track).
The beads that form the tetrahedral cage in FTC and ETC are linked with harmonic bonds,
\begin{equation}
    U_\mathrm{harmonic} = \frac{1}{2}k_{ij}|\mathbf{r}_{ij}|^2,
\label{eq:harmonic}
\end{equation}
where \(\mathbf{r}_{ij}\) is the distance vector between particle \(i\) and \(j\), and \(k_{ij}\) is the harmonic potential constant between particle \(i\) and \(j\). A C particle forms no bonds with the tetrahedron but is metastably trapped in the shell by steric repulsions. 
Decomposition of FTC into ETC + C occurs without breaking bonds.
For the shuttling and the track, the neighboring beads are linked with finitely extensible nonlinear elastic (FENE) bonds to prevent unrealistic bond stretching,
\begin{equation}
    U_\mathrm{FENE} = \frac{1}{2}\kappa_{ij}r_{0,ij}^2 \ln \Big[ 1 - \left( \frac{|\mathbf{r}_{ij}|}{r_{0,ij}}\right)^2\Big],
    \label{eq:FENE}
\end{equation}
where \(\kappa_{ij}\) is the FENE force constant between particles \(i\) and \(j\), \(r_{0,ij}\) is the maximum extension, and \(\mathbf{r}_{ij}\) is the distance vector.
The ring particles are further connected with angular potentials that help maintain the circular shape,
\begin{equation}
    U_\mathrm{angle} = \frac{1}{2}l_{ijk} (\theta_{ijk} - \theta_{0,ijk})^2
\label{eq:angular}
\end{equation}
where \(l_{ijk}\) is the angular potential constant, \(\theta_{ijk}\) is the angle formed by the three particles \(i\), \(j\) and \(k\) and \(\theta_{0,ijk}\) is the equilibrium angle.

To accelerate the simulation, we used a cell list, where the simulation box was divided into cells with dimensions no smaller than 4.25 on each dimension.
Correspondingly, the long-range LJ interactions switched smoothly to 0 between \(4.0 \leq |\mathbf{r}_{ij}| \leq 4.25\) to ensure the interactions were not abruptly cut off when crossing cells.
The cell-list LJ potential is,
\begin{equation}
U_{LJ}^\mathrm{cell}(\mathbf{r}_{ij}) =
\begin{cases}
U_{LJ}(\mathbf{r}_{ij}) & \text{if } |\mathbf{r}_{ij}| \leq 4.0\\
U_{LJ}(\mathbf{r}_{ij}) (1+\lambda^2(2\lambda-3)) & \text{if } 4.0 < |\mathbf{r}_{ij}| \leq 4.25\\
0 & \text{if } |\mathbf{r}_{ij}| > 4.25,
\end{cases}
\label{eq:modifiedLJ}
\end{equation}
with smoothing parameters $\lambda = (|\mathbf{r}_{ij}|-r_\mathrm{cut}+r_\mathrm{switch})/r_\mathrm{switch}$, $r_\mathrm{cut} = 4.25$, and $r_\mathrm{switch} = 0.25$.
Complete force-field parameters can be found in Table~\ref{tab:table1},~\ref{tab:table2}, and~\ref{tab:table3}.

\begin{table*}[h!]
\tiny
\centering
\begin{tabular}{|c|c|c|}
\hline
\bf{Particle Type} & \(m_{i}\) & \(\sigma_{i}\) \\
\hline
\hline
BIND & 1.0 & 1.0 \\
INERT & 1.0 & 1.0 \\
CAT1 & 1.0 & 1.0 \\
CAT2 & 1.0 & 1.0 \\
SHUTTLE & 1.0 & 1.0 \\
TET1 & 1.0 & 1.0 \\
TET2 & 1.0 & 1.0 \\
TET3 & 1.0 & 1.0 \\
TET4 & 1.0 & 1.0 \\
CENT & 1.0 & 0.45 \\
\hline
\end{tabular}
\caption{All particles are assigned unit radius and unit mass except for the central particle, which has a smaller radius.}
\label{tab:table1}
\end{table*}

\begin{table*}[h!]
\tiny
\centering
\begin{tabular}{|c|c|c|c|c|c|}
\hline
\bf{Interaction} & \(\epsilon_{\mathrm{R},ij}\) & \(\epsilon_{\mathrm{A},ij}\) & \bf{Bond Type} & \(k_{ij}|\) \(\kappa_{ij}\) & \(r_{\mathrm{max},ij}\) \\
\hline
\hline
TET*-TET* & 10.0 & 0.0 & harmonic & 120.0 & - \\
INERT-INERT & 1.0 & 0.0 & FENE & 30.0 & 1.5   \\
INERT-BIND & 1.0 & 0.0  & FENE & 30.0 & 1.5  \\
BIND-CAT2 & 1.0 & 0.0 & FENE & 30.0 & 1.5  \\
CAT2-CAT1 & 1.0 & 0.0 & FENE & 30.0 & 1.5  \\
CAT2-INERT & 1.0 & 0.0 & FENE & 30.0 & 1.5   \\
SHUTTLE-SHUTTLE & 1.0 & 0.0 & FENE & 30.0 & 1.5  \\
\hline
SHUTTLE-CAT1 & 1.0 & 0.0 & none & - & - \\
SHUTTLE-CAT2 & 1.0 & 0.0 & none & - & -  \\
SHUTTLE-TET* & 500000.0 & 0.0 & none & - & - \\
SHUTTLE-CENT & 100000.0 & 0.0 & none & - & -  \\
SHUTTLE-INERT & 1.0 & 0.0 & none & - & - \\
INERT-TET* & 1.0 & 0.0 & none & - & -  \\
INERT-CENT & 1.0 & 0.0 & none & - & -  \\
INERT-CAT1 & 1.0 & 0.0 & none & - & -   \\
INERT-CAT2 & 1.0 & 0.0 & none & - & -   \\
BIND-BIND & 1.0 & 0.0 & none & - & - \\
BIND-CAT1 & 1.0 & 0.0 & none & - & -  \\
BIND-CAT2 & 1.0 & 0.0 & none & - & -  \\
BIND-TET* & 1.0 & 0.0 & none & - & - \\
BIND-CENT & 1.0 & 0.0 & none & - & - \\
CAT1-CAT1 & 1.0 & 0.0 & none & - & -   \\
CAT2-CAT2 & 1.0 & 0.0 & none & - & -  \\
TET*-CENT & 1.0 & 0.0 & none & - & - \\
CENT-CENT & 100000.0 & 0.0 & none & - & - \\
\hline
BIND-SHUTTLE & 1.0 & {\bf 0.4 $\vert$ 0.7 $\vert$ 0.9} & none & - & - \\
CAT1-TET1 & 0.888468 & 0.732877 & none & - & - \\
CAT1-TET2 & 1.215096 & 0.573471 & none & - & - \\
CAT1-TET3 & 1.424995 & 0.221362  & none & - & - \\
CAT1-TET4 & 0.216518 & 0.511311  & none & - & - \\
CAT1-CENT & 2.798031 & 3.922903 & none & - & - \\
CAT2-TET1 & 1.147790 & 1.457497 & none & - & - \\
CAT2-TET2 & 1.440685 & 1.253230 & none & - & - \\
CAT2-TET3 & 2.063941 & 0.928847 & none & - & - \\
CAT2-TET4 & 3.788327 & 1.567248  & none & - & - \\
CAT2-CENT & 1.421851 & 1.196286 & none & - & - \\
\hline
\end{tabular}
\caption{
Pairwise interaction parameters for modified Lennard–Jones and bonded interactions.
Tetrahedral vertices are connected by harmonic bonds, while neighboring particles along the shuttling ring and track are linked by FENE bonds in the sequence BIND–CAT2–CAT1–CAT2–(INERT)$_8$, repeated four times.
Harmonic bond strengths \(k_{ij}\) follow Eq.~\ref{eq:harmonic}, and FENE bond strengths \(\kappa_{ij}\) follow Eq.~\ref{eq:FENE}.
An asterisk in TET* denotes identical values for TET1, TET2, TET3, and TET4.
}
\label{tab:table2}
\end{table*}

\begin{table*}[h!]
\tiny
\centering
\begin{tabular}{|c|c|c|c|}
\hline
\bf{Interaction} & \(l_{ijk}\) & \(\theta_{0,ijk}\) \\
\hline
\hline
SHUTTLE-SHUTTLE-SHUTTLE & 3.0 &\(  \pi \left(1 - \frac{2}{N_{\mathrm{shuttle}}}\right) \) \\

BIND-CAT2-CAT1 & 10.0 &\(  \pi \left(1 - \frac{2}{N_{\mathrm{large}}}\right) \) \\
CAT2-CAT1-CAT2 & 10.0 &\(  \pi \left(1 - \frac{2}{N_{\mathrm{large}}}\right) \) \\
CAT1-CAT2-INERT & 10.0 & \( \pi \left(1 - \frac{2}{N_{\mathrm{large}}}\right) \) \\
CAT2-INERT-INERT & 10.0 &\(  \pi \left(1 - \frac{2}{N_{\mathrm{large}}}\right) \) \\
INERT-INERT-INERT & 10.0 & \( \pi \left(1 - \frac{2}{N_{\mathrm{large}}}\right) \) \\
INERT-INERT-BIND & 10.0 & \(  \pi \left(1 - \frac{2}{N_{\mathrm{large}}}\right)  \) \\
INERT-BIND-CAT2 & 10.0 & \( \pi \left(1 - \frac{2}{N_{\mathrm{large}}}\right) \)  \\

\hline
\end{tabular}
\caption{Angular interaction parameters for Eq.~\ref{eq:angular}. 
Note that the angular interactions are symmetric about the center particle such that \(l_{ijk}\) would be equivalent to \(l_{kji}\).
}
\label{tab:table3}
\end{table*}
\subsection{Maintaining Nonequilibrium Steady-State}
To maintain a chemical potential gradient between fuel and waste species, we performed grand canonical Monte Carlo (GCMC) moves 
that add and remove FTC, ETC, and C into the outer region of the box.
These molecules were added and removed far enough from the motor that they were not directly interacting.
Simulations executed 100 time steps of Langevin dynamics followed by an attempted GCMC move from configuration \(x\) to \(x^\prime\) which was accepted with probability,
\begin{equation}
    P_\mathrm{acc}(x\rightarrow x^\prime) = \min\Big[1, \frac{\pi(x^\prime) P_\mathrm{gen}(x|x^\prime)}{\pi(x)P_\mathrm{gen}(x^\prime|x)} \Big],
\end{equation}
where \(\pi\) is the target distribution and \(P_\mathrm{gen}\) is the generation probability for an attempted move.
In the grand canonical ensemble, the target distribution \(\pi\) over the positions \(\mathbf{r}\), momenta \(\mathbf{p}\), and number \(N_i\) of each species \(i\) in the simulation box is,
\begin{equation}
    \pi(\mathbf{r},\mathbf{p},N) \propto  \frac{1}{\prod\limits_i N_i!}
    e^{-\beta [U(\mathbf{r})+K(\mathbf{p})-\sum\limits_i \mu_iN_i]},
\end{equation}
where \(U\) and \(K\) are the potential and kinetic energies, and \(\mu_i\) is the chemical potential of species \(i\).

We uniformly picked a species from FTC, ETC and C and decided to attempt deletion or addition.
The momenta of the species were drawn from the Boltzmann distribution and we had pre-sampled libraries for FTC and ETC that contain \(10^4\) configurations in the canonical distribution.
Therefore, for each species \(i\), the generation probability for an insertion is,
\begin{equation}
    P_\mathrm{gen}(N_{i} \rightarrow N_i+1) = \frac{1}{N_i +1}\frac{e^{-\beta H_0}}{Z_i^0},
\end{equation}
and the deletion probability (reversed probability, removing the same molecule) is
\begin{equation}
    P_\mathrm{gen}( N_i+1 \rightarrow N_{i}) = \frac{1}{N_i +1}.
\end{equation}
Here \(H_0 = U_0+K_0\) is the total internal energy of the inserted species with \(U_0\) the potential energy and \(K_0\) the kinetic energy, and \(Z_i^0\) is the single-molecule partition function in the box volume.
Plugging into the Metropolis criterion yields,
\begin{equation}
    P_\mathrm{acc}(N_i \rightarrow N_i+1) = \min\Big [1, \frac{1}{N_i+1} e^{-\beta (\Delta U -\mu_i)}Z_i^0\Big ],
\end{equation}
where \(\Delta U\) is the energy difference before and after insertion, the solvation energy of the inserted species. 
Defining the modified chemical potentials \(\mu_i^\prime = \mu_i - \log Z_i^0/\beta\) simplifies the expression,
\begin{equation}
    P_\mathrm{acc}(N_i \rightarrow N_i+1) = \min\Big [1, \frac{1}{N_i+1} e^{-\beta (\Delta U -\mu_i^\prime)}\Big ],
\end{equation}
and similarly, we obtained the acceptance ratio of the deletion attempts,
\begin{equation}
    P_\mathrm{acc}(N_i +1\rightarrow N_i) = \min \Big[ 1,(N_i +1) e^{\beta (\Delta U-\mu_i^\prime)}\Big].
\end{equation}
In the simulation, we set relative chemical potentials \(\mu_\mathrm{FTC}^\prime = 0.5\), \(\mu_\mathrm{ETC}^\prime = -10\), and \(\mu_\mathrm{C}^\prime = -10\) to maintain a FTC around 10 per box and no free ETC and C in the outer box.
\section{Important Timescales in the Nonequilibrium Molecular Dynamics Simulations}
While trajectories are generated with an underdamped Langevin integrator, we have previously shown that the \(\gamma = 0.5\) damping coefficient corresponds to an overdamped motor.
That analysis is, found in Sec.~3 of the SI to Ref.~\cite{Albaugh_2023}, highlights that different fundamental steps of the motor operation occur on vastly different timescales.
We reproduce those timescales here as they are essential to the mechanism.
Namely, C diffusion is very fast.
Diffusion of FTC and ETC as well as \(\mathrm{FTC\rightarrow ETC +C}\) reactions are almost 100 times slower.
Ring diffusion is yet another order of magnitude slower, which is why a current boost requires one to enhance the activity.

\begin{table*}[h!]
\centering
\caption{Timescales for fundamental processes with damping coefficient \(\gamma = 0.5\).}
\label{tab:timescale}
\begin{tabular}{ | c | c |}
\hline
C diffusion & \(6 \times 10^{-2}\) \\
damping & \(2 \times 10^0\) \\
ETC diffusion & \(2 \times 10^{0}\) \\
FTC diffusion & \(3 \times 10^{0}\) \\
catalyzed FTC decomposition & \(3 \times 10^{2}\) \\
ring diffusion & \(1 \times 10^{3}\) \\
current & \(8 \times 10^{3}\) \\
\hline
\end{tabular}
\label{tab:table4}
\end{table*}

\begin{figure}[h!]
\centering
\includegraphics[width=0.8\textwidth]{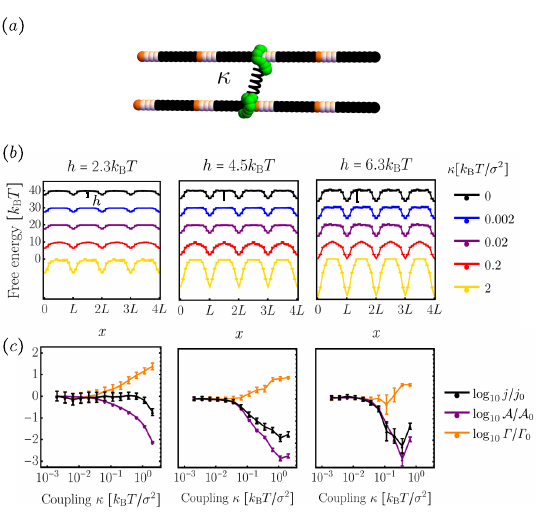}
\caption{
\textbf{Aligned configuration deepens the free energy and never boosts current.}
\textbf{(a)} Schematics of motors coupled in aligned-track configuration with coupling strength \(\kappa\).
\textbf{(b)} By tuning the pair potential between green shuttling ring particles and orange binding site particles, deeper wells can be induced.
Columns, labeled by the binding-well depth \(h\) for the uncoupled motor (black line), show how these wells grow monotonically deeper with increasing coupling \(\kappa\).
\textbf{(c)} Higher energy barriers prolong the escape time from the wells, leading to a monotonic decrease in activity with increasing \(\kappa\).
Although the bias increases, it does not compensate for the loss in activity. 
}
\label{fig:figS1}
\end{figure}
\section{Analysis of Equilibrium Diffusion on Aligned Tracks}
Green shuttling rings bind to orange binding sites due to the attractive term in the LJ potential (Eq.~5 in the main text) with the attraction strength $\epsilon_{\mathrm{A}, \mathrm{BIND-SHUTTLE}}$, which tunes the depth of the binding well for uncoupled motors.
When no FTC, ETC, or C are included in the simulations, the system is in equilibrium.
Under those conditions, free energy calculations (see SI Sec.~5) allow us to construct a 1D free energy indicating the propensity for the center of mass of the ring to be positioned at different locations along the track.
We label the different $\epsilon_{\mathrm{A}, \mathrm{BIND-SHUTTLE}}$ values by the depth of that free energy well $h$ in the case of equilibrium uncoupled motors.
When the two rings are uncoupled, \(\epsilon_{\mathrm{A}, \mathrm{BIND-SHUTTLE}} = 0.8, 1.4 \textrm{,\,and\,} 1.8 k_\mathrm{B}T\) produce well depth \(h = 2.3, 4.5\textrm{,\,and\,}  6.3 k_\mathrm{B}T\), corresponding to each column of Fig.~\ref{fig:figS1} (a) and (b) respectively.

When the two rings are coupled in the aligned configuration (Fig.~\ref{fig:figS1} (a)), an attempted escape by one ring is held back by the other.
Increasing the coupling \(\kappa\) deepens the free energy wells, as shown in Fig.~\ref{fig:figS1} (b).
The increased barrier height causes a monotonic decrease in activity across all binding well depths shown in Fig.~\ref{fig:figS1} (c).
The longer timescale for escaping has an opposing effect on bias.
As discussed in the main text, longer time spent on the binding site increases kinetic asymmetry, resulting in a larger bias.

\section{Current Boosts via Explicit-Linker Coupling}
\begin{figure}[h!]
\centering
\includegraphics[width=0.8\textwidth]{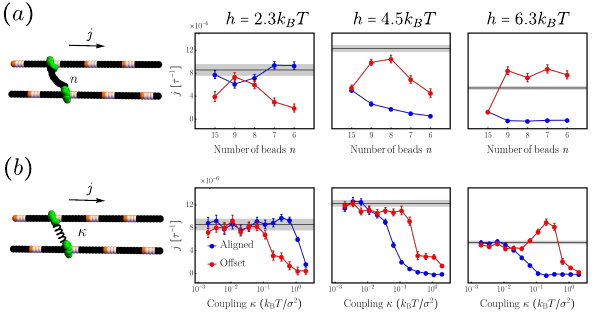}
\caption{
\textbf{Coupling via explicit linkers yields a current boost under similar conditions as spring-coupled motors.}
\textbf{(a)} 
The two motors are coupled via explicit linkers with \(n\) intermediate particles.
Shorter linkers generate stronger coupling forces.
Current boosts are only observed for offset-track configuration, shorter linkers (\(n \leq 9\)), and \(h = 6.3 k_\mathrm{B}T\).
\textbf{(b)} 
To highlight the similarity in boost conditions, currents from spring-coupled motors under matched conditions are shown for comparison. 
These data are compiled from the Fig.~\ref{fig:fig1},~\ref{fig:fig2} and~\ref{fig:fig3} of main paper. 
Means and error bars are collected from 100 independent simulations for each motor.
}
\label{fig:figS2}
\end{figure}
In Fig.~1 of the main text, we confirmed that motors coupled harmonically can yield more currents compared to uncoupled motors.
Here, we extended the study to coupling with explicit linkers as shown in Fig.~\ref{fig:figS2}.
That nonlinear coupling force is achieved by connecting two particles from each ring via a chain consisting of \(n\) intermediate particles, where neighboring particles interact through a finitely extensible nonlinear elastic (FENE) potential, the same potential used to hold particles together in the shuttling ring (Eq.~\ref{eq:FENE}). 
The same simulation conditions and forcefield parameters used for spring coupling are applied to explicit-linker simulations. 
The current enhancements persist under similar structural, energetic, and coupling strengths, demonstrating the robustness of the boosting effect.
\section{Equilibrium Free Energy Calculations}
\label{sec:FE}
\begin{figure}[h!]
\centering
\includegraphics[width=0.8\textwidth]{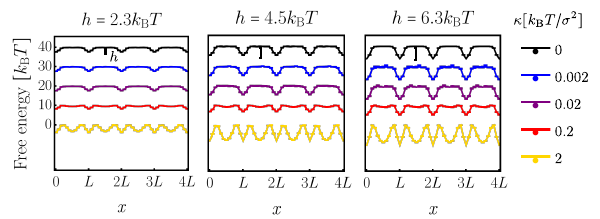}
\caption{
\textbf{Free energy profile in the offset configuration at equilibrium.}
For a single-track motor (black), the ring's activity is regulated by the free energy barrier \(h\) required to climb out of the binding site.
Coupling two motors together make the barriers significantly shallower at intermediate coupling strength \(\kappa = 0.2 k_\mathrm{B}T/\sigma^2\) (red) when \(h = 4.5\) and \(6.3 k_\mathrm{B}T\).
When the spring is too strong, \(\kappa = 2 k_\mathrm{B}T/\sigma^2\)  (yellow), the state with the ring between two binding sites becomes metastable.
The cartoon picture in Fig.~\ref{fig:fig2} of main text uses the data at \(h = 6.3 k_\mathrm{B}T\), and \(\kappa = 0 \text{ and } 0.2 k_\mathrm{B}T/\sigma^2\).
}
\label{fig:figS3}
\end{figure}
To explain the activity boost, we have analyzed how the unfueled shuttling ring's potential of mean force depends on coupling \(\kappa\), attraction strength \(\epsilon_{\mathrm{A}, \mathrm{BIND-SHUTTLE}}\), and the alignment of the tracks.
For conditions with comparatively shallow wells, it was sufficient to collect 100 independent unbiased equilibrium trajectories, each \(2\times 10^8\) time step long, for motors under various coupling strengths with no FTC, ETC, or C.
The tracks have 12 particles before they periodically repeat.
We averaged over all times, both tracks, and repeated units to collect \(P(i)\), the probability that a ring's center of mass (COM) would be closest to \(i=1,2,\cdots,12\).
Defining the orange binding site to be \(i =1\) and placing that particle at \(x=0\), we transform to \(P(x)\) and plot free energy \(-\ln P(x)\), equivalently the potential of mean force felt by the COM of a shuttling ring.

The free energies for aligned and offset tracks are plotted in Fig.~\ref{fig:figS1} and~\ref{fig:figS3}, respectively.
The attractive LJ parameter \(\epsilon_{\mathrm{A}, \mathrm{BIND-SHUTTLE}}\) between the orange and green particles was set to 0.8, 1.4 and 1.8 \(k_\mathrm{B} T\), with each value corresponding to another column of the figures.
Because the ring consists of multiple green particles, it is not intuitively obvious how tightly the ring sticks to the binding site for a particular value of $\epsilon_{\mathrm{A}, \mathrm{BIND-SHUTTLE}}$.
We find it more natural to characterize that strength by \(h\), the depth of the \emph{uncoupled} free energy well, measured from the black line in Fig.~\ref{fig:figS1} or~\ref{fig:figS3}.
Depending on track alignment, those wells could grow deeper or shallower as \(\kappa\) was increased.

At intermediate coupling strength (red in Figs.~\ref{fig:figS1} and~\ref{fig:figS3}) and offset configuration, the coupling significantly lowers the free energy to escape from one of the deeper barriers (\(h = 4.5\) and \(6.3 k_\mathrm{B}T\)), but coupling does not provide a similar benefit for the shallow well  (\(h = 2.3k_\mathrm{B}T\)).
In the cases where coupling lowers the free energy barrier, ring hopping between binding sites is facilitated, thereby enhancing the activity in line with the mechanism shown in Fig.~\ref{fig:fig2} in the main text.
Further increasing the coupling strength (yellow) stabilizes intermediate states (with shuttling ring between two binding sites).
The ring also starts to deform and ``pinch'' the track, which leads to an even deeper well than the uncoupled motors.

For cases where sampling remained insufficient (i.e., strong spring strength, \(\kappa = 2k_\mathrm{B}T/\sigma^2\)), we applied umbrella sampling to enhance sampling across \(x\) and used the multistate Bennett acceptance ratio (MBAR) \cite{Shirts_2008} to reconstruct the unbiased free-energy landscape.
Harmonic biasing potentials were applied to the COM of the ring, \(U_\mathrm{biased} = \frac{1}{2}k(x-x_0)^2\), where \(k\) is the spring strength of the biased potential, and \(x_0\) is the center.
Each biased potential was evaluated from 10 trajectories with \(5\times 10^6\) time steps.
Four sets of biased potentials were used: a. \(k =0,\, x_0=0\) (uncoupled), b. \(k = 1, x_0 = 0,1,2,\cdots,11\), c. \(k = 5, x_0 = 0,1,2,\cdots,11\), d. \(k = 1 \text{ and } 5, x_0 = 2.5 \text{ and } 9.5\), in the interval \([0,12)\).
The resulting free energy profile within the primary interval was then replicated across the other units of the track to complete the four periodic landscapes.
\section{Parameterization of a Jump-Diffusion Model} 
\begin{figure}[h!]
\centering
\includegraphics[width=0.7\textwidth]{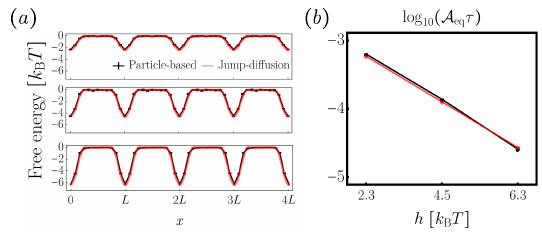}
\caption{
\textbf{Jump-diffusion model at Equilibrium.}
(a) Fitting the bare-track potential to free energy landscape of COM of the ring in MD simulations with \(h\) = 2.3, 4.5 and 6.3 \(k_\mathrm{B}T\), and \(w = 6\).
(b) The equilibrium activities \(\mathcal{A}_{eq}\) are computed from single-track simulations with no fuel present, both in the MD (black) and JD (red) simulations.
}
\label{fig:figS4}
\end{figure}

The jump-diffusion (JD) model was parameterized to recapitulate the essential physics of the MD model.
First, we parameterized the bare-track potential by fitting an analytical form (Eq.~7) to the free energy landscapes obtained from MD. 
With well width \(w = 6\) and barrier height \(h = 2.3, 4.5 \text{ and } 6.3 k_\mathrm{B}T\) measured from MD, the JD potential captures the overall shape and scale of the MD free energy profiles shown in Fig.~\ref{fig:figS4}(a).
Fig.~\ref{fig:figS4}(b) shows that the equilibrium activity of an unfueled ring is consistent between MD and JD.

\begin{figure}[h!]
\centering
\includegraphics[width=0.7\textwidth]{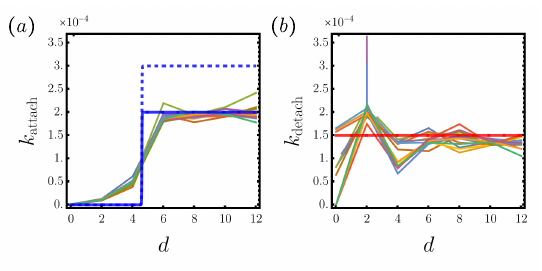}
\caption{
\textbf{The distance dependence of the rates of attaching and detaching a blocking group.}
(a) The \(k_\mathrm{attach}\) in the MD simulation are calculated as a function of the distance \(d\) between COM of the ring and each catalytic sites (thin, colored) from simulations at different conditions (various coupling strengths, and phase shifts). 
Each thin colored line represents statistics averaged over 100 simulations. 
In the JD model, the attachment rate \(k_\mathrm{attach}(d)\) is fitted with a Fermi function (blue, solid), with a right-tail value of \(k_\mathrm{attach}^\mathrm{far} = 2 \times 10^{-4}\). 
The dashed blue line presents an example of variations on \(k_\mathrm{attach}^\mathrm{far}\) realized in the JD model.
The center of the fermi function is 4.6 for the rate to smoothly transition from 0 to \(k_\mathrm{attach}^\mathrm{far}\).
(b) The detachment rate on the far side is approximately an constant, with \(k_\mathrm{detach} = 1.5 \times 10^{-4}\) (red line), fixed in the JD model.
}
\label{fig:figS5}
\end{figure}

To enforce the chemical rate rules that have been seen in the MD, we collected the attachment and detachment rates from MD trajectories under various simulation conditions in Fig.~\ref{fig:figS5}.
Fig.~\ref{fig:figS5}(a) shows a converged trend that the attachment rate increases smoothly with ring-blocking-group distance \(d\) and can be well represented by a Fermi function.
The large-\(d\) limit, \(k_\mathrm{attach}^\mathrm{far}= 2\times 10^{-4}\), sets the baseline attachment rate far from the catalytic site.
The midpoint, near \(d= 4.6\), corresponds to the steric radius at which attachment becomes accessible and sets the center of Fermi functions.
We fixed the sharpness of the transition in \(k_\mathrm{attach}^\mathrm{far}\) and made it sharper (i.e., a steeper Fermi function) than in MD model intentionally.
If we choose the broader MD width while increasing \(k_\mathrm{attach}^\mathrm{far}\) across orders of magnitudes, \(k_\mathrm{attach}^\mathrm{far}\) would become non-negligible even at extremely small separations between the ring and the catalytic site.
This would allow attachment of the blocking group even when the ring is effectively on top of the catalytic site.
That behavior violates the steric exclusion rule built into the microscopic MD model.
Detachment events are less distance-dependent and modeled as a constant \(k_\mathrm{detach} = 1.5 \times 10^{-4} \) (Fig.~\ref{fig:figS5}(b)).
We varied \(k_\mathrm{attach}^\mathrm{far}\) to study the impact of the concentration of fuels.

\section{Coupling Motor Extends Boosts Beyond Fueling Single-Motor Limits}
\begin{figure}[ht]
\centering
\includegraphics[width=0.8\textwidth]{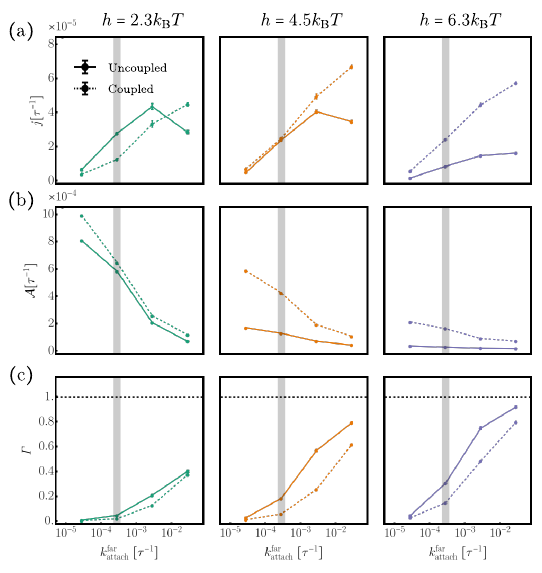}
\caption{
\textbf{
Current, Activity, and Bias for Coupled and Uncoupled Motors.
  }
  The JD model provides access to higher fuel.
\textbf{(a)} The current \(j\) is saturated at high \(k_\mathrm{attach}^\mathrm{far}\) for the uncoupled motor.
Coupling is able to provide additional current even when fuel saturates.
The gray line denotes the \(k_\mathrm{attach}^\mathrm{far}\) value fitted from MD simulations.
\textbf{(b)} 
For the uncoupled motor, activity \(\mathcal{A}\) decreases as fuel increases due to longer occupation of blocking groups, whereas motor coupling enhances activity and alleviates this kinetic limitation.
\textbf{(c)} In the uncoupled case, the bias \(\mathit{\Gamma}\) increases with \(k_\mathrm{attach}^\mathrm{far}\), indicating a stronger directionality but less frequent transition. 
Coupling reduces the bias while maintaining a sufficient activity, enabling the coupled system to exceed the maximal current attainable by a single motor at high fuel concentration.
The coupled values relative to uncoupled values are shown in Fig.~\ref{fig:fig4} of the main text.
}
\label{fig:figS6}
\end{figure}
One way to increase the current is to raise the chemical driving force by increasing the fuel concentration. However, for a single (uncoupled) motor, the current saturates at high fuel concentration, as shown in Fig.~\ref{fig:figS6}(a), such that further increases in fuel do not lead to additional current.

The activity–bias decomposition in Fig.~\ref{fig:figS6}(b) and (c) reveals the origin of this saturation. In the high-fuel regime, the bias of the uncoupled motor approaches 1, indicating highly directional transitions. 
At the same time, the catalytic sites are occupied for most of the time, and the current becomes limited by ring's hopping between binding sites. 
As a result, the current is constrained by reduced activity rather than by insufficient driving force.

Coupling two motors qualitatively changes this behavior. While coupling reduces the bias, it significantly enhances the overall activity of the motor complex, allowing transitions to occur more frequently. 
This reallocation between bias and activity alleviates the kinetic bottleneck responsible for saturation in the uncoupled case, enabling the coupled system to exceed the maximal current achievable by a single motor even under high-fuel conditions.
\section{Sensitivity to JD Model Implementation}
\begin{figure}[ht]
\centering
\includegraphics[width=0.35\textwidth]{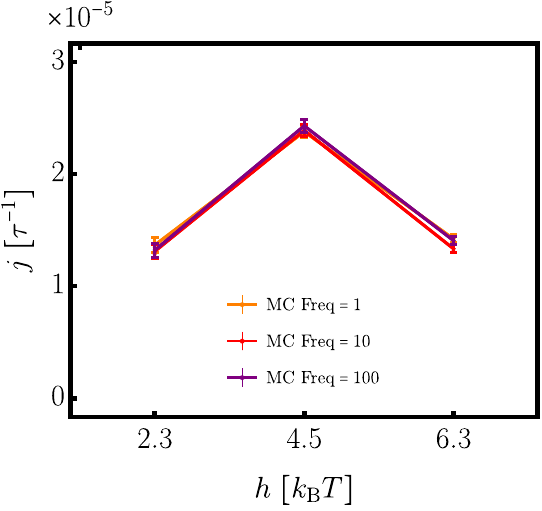}
\caption{
\textbf{Convergence of Monte Carlo attempt frequency in the jump–diffusion model.}
Current for uncoupled rings at \(k_\mathrm{attach}^\mathrm{far}=0.02\) is shown for Monte Carlo attempts every 1, 10, and 100 Langevin steps across well depths \(h\). 
The results overlap within error, indicating that an interval of 100 steps is sufficient; in this case the probability for an event happen is \(k(100\Delta t) \leq 1\%\).
}
\label{fig:figS7}
\end{figure}

\begin{figure}[ht]
\centering
\includegraphics[width=0.7\textwidth]{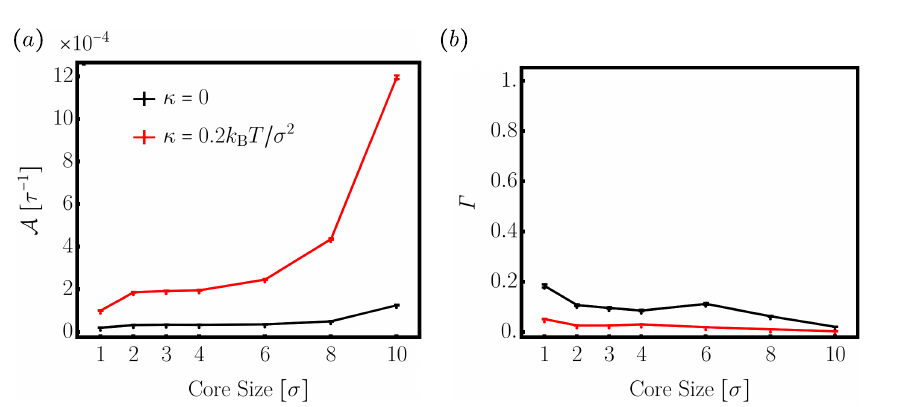}
\caption{
\textbf{Effect of core size on current decomposition into activity and bias.}
(a) Activity \(\mathcal{A}\) for uncoupled and coupled motors at \(k_\mathrm{attach}^\mathrm{far} = 2\times 10^{-4}\) are evaluated with different choices of the core region used to define hops.
The coupled motor is more sensitive to the core size, but both cases show a plateau when the core size is between 2\(\sigma\) and 4\(\sigma\), indicating stable counting. Too small a core underestimates activity, whereas too large a core overestimates it.
(b) The same plateau region is observed for the bias \(\mathit{\Gamma}\).
A core size of 3\(\sigma\) is chosen throughout the paper to count hopping events.
}
\label{fig:figS8}
\end{figure}

In the JD model, chemical events are sampled stochastically by Monte Carlo attempts performed at fixed intervals of the Langevin integrations. 
To ensure the choice of intervals does not cause a periodic pumping effect and does not bias the results, we varied the attempt frequency from every 1 to every 100 Langevin steps.
As shown in Fig.~\ref{fig:figS7}, the currents overlap within statistical error across different intervals, confirming that an interval of 100 steps is sufficient. 
This test was performed at the largest \(k_\mathrm{attach}^\mathrm{far}\), where the probability of attempted events is maximized but still less than \(1\%\).
To ensure that the current decomposition into activity and bias is not biased by the definition of hopping events, we systematically varied the core size and identified a plateau region in which both activity and bias were stable (Fig.~\ref{fig:figS8}). 
If the core size is chosen too small, hops are undercounted because dwells beyond the binding well tips are missed. 
If it is too large, hops are overcounted because random diffusion around the boundary are misclassified as completed transitions. 
Within the plateau region, it yields robust counts.
For the JD model, a diameter of \(3\sigma\) (half well width) was chosen.
Though MD model adopted a core size of \(\sigma\) (the size of binding site), all qualitative trends and conclusions from the MD study remain unaffected.
\end{document}